\documentclass[aip,jap]{revtex4-1}
\usepackage{graphicx}
\usepackage{dcolumn}
\usepackage{bm}

\begin{document}

\title{Design and fabrication of robust broadband extreme ultraviolet multilayers} 



\author{Shang-qi Kuang}
\email{physicskuang@sina.com}
\author{Jian-bo Wang}
\affiliation{School of Science, Changchun University of Science and Technology, Changchun 130022, China
}
\author{Hai-gui Yang}
\affiliation{Adavanced Manufacturing Technology for Optical Systems Laboratory, Changchun Institute of Optics, Fine Mechanics and Physics, Chinese Academy of Sciences, Changchun 130033, China}

\author{Tong-lin Huo}
\author{Hong-jun Zhou}
\affiliation{Synchrotron Radiation Laboratory, University of Science and Technology of China, Hefei 230029, China}



\date{\today}

\begin{abstract}
The random layer thickness variations can induce a great deformation of the experimental reflection of broadband extreme ultraviolet multilayer. In order to reduce this influence of random layer thickness fluctuations, the multiobjective genetic algorithm has been improved and used in the robust design of multilayer with a broad angular bandpass. The robust multilayer with a lower sensitivity to random thickness errors have been obtained and the corresponding multilayer mirrors were fabricated. The experimental results of robust Mo/Si multilayer with a wide angular band were presented and analyzed, and the advantage of robust multilayer design was demonstrated.  
\end{abstract}


\maketitle 

\section{Introduction}
Due to the great improvement of extreme ultraviolet (EUV) lithography for the semiconductor industry, these reseraches of
EUV multilayers or multialyer interference coatings attract many attention \cite{Louis:2011,Glatzel:2013,Huang:2017}. According to the Bragg condition, the EUV multilayers have a periodic layer structure, and the thickness of each layer is a few nanometers. Due to the inherently limitation induced by the saturated number of bilayers, the periodic multilayers only can supply the high reflectivity in the narrow incidence range at a specific wavelength \cite{Medvedev:2013,Kuhlmann:2002,Yulin:2003}. Therefore, the applications of periodic multilayers are limited for optical systems such as EUV lithography \cite{Yakshin:2010}, astronomy telescope \cite{Yao:2013}, and soft X-ray microscopy, especially for the optical system with a high numerical aperture (NA) \cite{Huang:2017,Yakshin:2010}. In order to increase the bandwidth of reflection peak, the multilayer design has been changed from the periodic multilayer structure to the aperiodic layer arrangement \cite{Kuhlmann:2002,Yulin:2003,Yao:2013,Yakshin:2010}. 

In the design of aperiodic multilayers with a wide bandwidth of incidence angle, several approaches in theory have been demostrated. The optimized apriodic multilayer structures can be obtained by these approaches, which are based on numerical optimization \cite{Cheng:2006,Wang:2006,Wang:2011} or a combination of analytical designing and numerical optimization \cite{Kozhevnikov:2001,Morawe:2002}. In these optimizations, the thicknesses of the layers have been considered as a set of independent variavles, and the solution is a set of optimized layer thicknesses which can provide a distinct minimum of the merit function. In this scheme, this calculated merit function is the deviation of the calculated reflectivity profile from the aimed one, and it is not very difficult to obtain a desirable layer thickness distribution. However, the theoretical simulations and experimetal results demonstrated that the natural interlayers and thickness errors can induce a great deformation of the reflectivy curve \cite{Yakshin:2010}. In order to keep the structure of interlayers the same over the whole multilayer, a design procedure which constrains the layer thickness variation has been given, and the desinged thickness variation does not exceed 0.39nm \cite{Kozhevnikov:2015}. Keeping the layer thickness in a small range, it is a good assumption to use the same properties for interlayer in the whole multilayer stack, but the influence of random thickness flunction was not considered. With the purpose of considering the random thickness errors in the process of broadband multilayer design, the multiobjective evlutionary algorithm \cite{Deb:2002} has been applied in the designing process, and the robust multilayer designs have been obtained theoretically \cite{Kuang:2018}.

In this paper, the robust design method of broadband EUV multilayer has been improved to adapt the control precision of layer thickness, and the designed Mo/Si multilayers with practical parameters have been obtained and demonstrated. According to the robust designs of Mo/Si multilayers, the EUV multilayer coatings have been fabricated and characterized, and these results present the feasibility and advantage of robust multilayer design based on multiobjective evolutionary algorithm in the researches of broadband EUV mirrors.  

\section{Robust multilayer design according to control precision of layer thickness}
In a realistic Mo/Si multilayer system, the reflectivity is sensitive to the imperfections of interface \cite{Braun:2014}, the interlayers \cite{Aquila:2006} and the oxidation of top layer \cite{Yakshin:2010}, and then all these effects have been considered in our simulations. Because the thicknesses of Mo and Si layers are contrained in a small range, it is a good assumption that these two interlayers have the same chemical composition and layer thicknesses over the whole stack. Therefore, the structure of multilayer system can be written as sub/[MoSi$_{2}$/Mo/MoSi$_{2}$/Si]$_{49}$/SiO$_{2}$, where the layers of MoSi$_{2}$ are the interlayers and SiO$_{2}$ oxide layer results from the oxidation of the top silicon layer. The parameters of the interlayers and SiO$_{2}$ oxide layer can be obtained by charactering the periodic Mo/Si multilayer \cite{Kuang:2017}. 

The details of theoretical calculations of the reflectivity of non-periodic multilayer system have been demonstrated \cite{Kuang:2018}, and these two merit functions of multilayer design can be given by \cite{Kuang:2018,Wu:2012}
	\begin{eqnarray}
	f_{1}&=&\int_{\theta_{\min}}^{\theta_{\max}}[R(\theta)-R_{0}]^{2}d\theta; \nonumber \\
	f_{2}&=&f_{1}+\frac{1}{2}\sum_{i=1}^{98}\frac{\partial^{2}f_{1}}{\partial d_{i}^{2}}\delta_{i}^{2},
	\end{eqnarray}
where $R(\theta)$ and $R_{0}$ are the calculated and aimed reflectivities of designed multilayers, respectively. The first merit function $f_{1}$ characterizes the root-mean-square deviation of calculated reflectivity profile from the desired one. In the first merit function, $\theta$ is the incidence angle, and the constant reflectivity $R_{0}$ is 52$\%$ in the range of incidence angle $[0^{\circ},16^{\circ}]$ at a wavelength of 13.5nm. The second function $f_{2}$ is the robust design merit function, which characterizes the sensitivity of the reflectivity curve of designed multilayer to the random fluctuations of layer thicknesses. In the second merit function, $d_{i}$ and $\delta_{i}$ are the thickness and  random thickness error's standard deviation of the $i$th layer respectively. Therefore, the designed aim is to optimize the layer thicknesses of Mo and Si layers to simultaneously minimize these two merit functions, and a multiobjective evolutioary algorithm has been applied in the optimized process \cite{Kuang:2018}. 

\begin{figure}
\includegraphics[width=10cm]{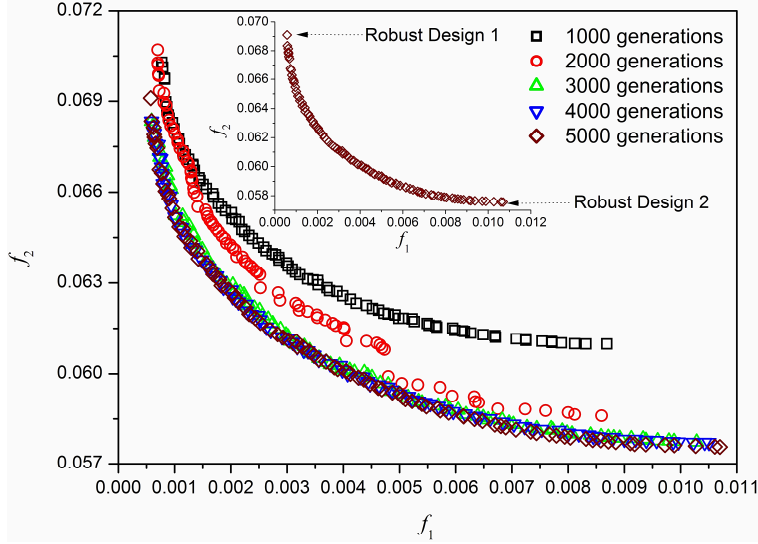}
 \caption{Obtained nondominated solutions according to different generation of multiobjective evolutionary algorithm, where the reflectivity and robust performances of Mo/Si multilayer with a wide angular bandpass are set as two optimized objectives. The design of aperiodic multilayer is optimized to obtain a constant reflectivity $R_{0}=52\%$ in the range of incidence angle $[0^{\circ},16^{\circ}]$ at a wavelength of 13.5nm, and here the S polarized radiation is considered. The inset shows the nondominated solutions for the 5000th generation, and these two boundary solutions in the nondominated front are named as ``Robust Design 1" and ``Robust Design 2", respectively. In the theoretical calculations, the random thickness error of Mo or Si layer has a normal distribution and the standard deviation of 0.15nm.}
\end{figure}

In the experiment, the designed aperiodic Mo/Si multilayer was deposited on a polished silicon substrate by a direct current magnetron sputtering system. The Mo and Si targets were operated at powers of 30W and 20W, respectively. The base pressure of vacuum system is $2\times10^{-4}$ Pa, and the deposition is performed under Ar atmosphere of $0.1$ Pa. The average deposition rates for Si and Mo were 0.32nm/s and 0.19nm/s, respectively, which were calculated from periodic Mo/Si multilayers with different average periodic thciknesses determined by X-ray reflectivity measurements \cite{LEE:2002}. In our scheme, the layer thickness is controlled by time and the time control precision is one second, thus the deposition rate means the control precision of the layer thickness. In order to obtain a multilayer design which consists of layer thicknesses controlled in a high precision, only a series of discrete thicknesses is accepted in the optimization based on multiobjective evolutioary algorithm. Due to the strong searching ability of this algorithm, one can obtain the required multilayer designs with an acceptable reflected profile. 

Considering the control precisions of layer thicknesses, for simplicity and without loss generality, both the thickness errors of Mo and Si layers are assumed to have a normal distribution and a standard deviation of 0.15nm. In Fig. 1, the nondominated solutions for designed Mo/Si multilayers with a wide angular bandpass are presented, and all the individuals local at the first nondominated front after 3000 generations. The inset of Fig. 1 demonstrates the nondominated solutions of the 5000th generation, it is revealed that the relation between the reflection and robust performances of multilayer design is restrictive. Furthermore, the solutions with best reflectivity profile and lowest sensitivity to random layer thickness errors are defined as ``Robust Design 1" and ``Robust Design 2", respectively.

\begin{figure}
\includegraphics[width=10cm]{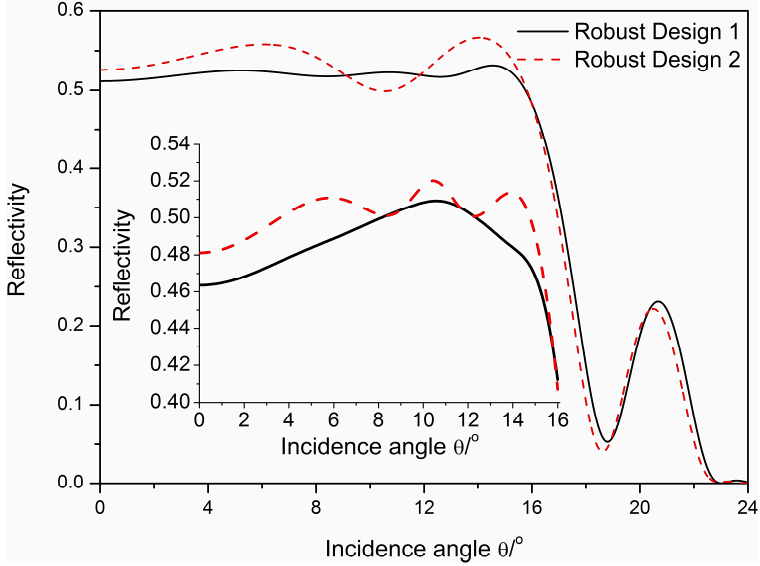}
\caption{Theoretical reflectances of robust designs of Mo/Si multilayer with a broad incidence angular of reflection, where these two robust designs are correspoding to boundary solutions in the nondominated front as shown as inset in Fig. 1. The inset shows the mathematical expectation reflectance of Robust Design 1 and Robust Design 2, respectively. In the theoretical simulations, the random thickness error of Mo or Si layer has the normal distribution and the standard deviation of 0.15nm. Here only the S polarized radiation is calculated.}
\end{figure}

The theoretical reflectivity plateaus of multilayer designs of Robust Design 1 and Robust Design 2 are demonstrated in Fig. 2, and here the random thickness fluctuations are not considered. After an investigation of Fig. 2, it is found that the reflectivity profile of Robust Design 1 is very close to the design target, while the reflectivity curve of Robust Design 2 has much more fluctuations. In order to analyse the influence of random layer thickness errors on the reflectivity plateau, the mathematical expectation of the reflectivity $\tilde{R}(\theta)$ should be analyzed, and it can be calculated by \cite{Kuang:2018,Furman:1992} 

\begin{equation}
\tilde{R}(\theta)=R(\theta)+\frac{1}{2}\sum_{i=1}^{98}\frac{\partial^{2}R(\theta)}{\partial d_{i}^{2}}\delta^{2}_{i}.
\end{equation}

In the inset of Fig. 2, the mathematical expectation reflectivity profiles of robust multilayer designs have been shown. It is presented that mathematical expectation of reflectivity curve of Robust Design 2 is higher than that of Robust Design 1, and it means that one has a higher probability to obtain a higher reflectivity plateau in experiments, if the multilayer design of Robust Design 2 is used. These results deonstrate that the reflectivity profile of multilayer design with a smaller value of the second merit function in Eq. (1) could be more stable than that of other multilayer designs. It is worthwhile to point out that the all the parameters used in our calculations are from the experiments, and these results demonstrate the strong adaptability of multiobjective evolutioanry algorithm in the multilayer design.

\section{Experimental results and analyses}
The measured reflectivity profiles of broadband EUV mirrors based on designed thickness distributions of Robust Design 1 and Robust Design 2 are demonstrated in Fig. 3, and these reflectivities were measured by the reflectometer at National Synchrotron Radiation Laboratory in Hefei, China. An investigation of Fig. 3 shows that the measured reflectivity curves have deformations compared with the designed results given in Fig. 2, no matter which multilayer design has been used. However, the measured reflectivity which varies between $45\%$ and $55\%$ in the $[0^{\circ},16^{\circ}]$ range of angle of incidence has been obtained. Comparing these two experimental results, it is found that at the small incidence angle, the observed reflectivity based on the multilayer design of Robust Design 2 is higher than that of Robust Design 1, and this result is consistent with the theoretical expectation of reflectivities as shown in the inset of Fig. 2. As shown in Fig. 4, the used thickness of Mo layer is in the range [1.75nm, 2.12nm], and the used range of the thickness of Si layer is [3.10nm, 4.37nm]. Therefore, the changed regions of layer thicknesses of Mo and Si layers are all very small, and it is a good approximation that the interfacial roughness and natural interlayers are fixed over the whole stack \cite{Aquila:2006,Yakshin:2000}. Hence, the deformations of measured reflectivity plateaus in Fig. 3 should be mainly induced by the random layer thickness errors arising during the deposition. In order to confirm this understanding, the experimental reflectivity curves are fitted by allowing the random thickness fluctuations in the controlling precision, and the fitting results are given in Fig. 3. An investigation of Fig. 3 demonstrates that the fitting results are well coincided with the experimental results. Although the thickness distribution obtained by fitting method should not be unique, but the good fitting results can allow us to make the conclusion that the deformation of the reflectivity curve is induced by the random layer thickness deviations.       

\begin{figure}
	\includegraphics[width=14cm]{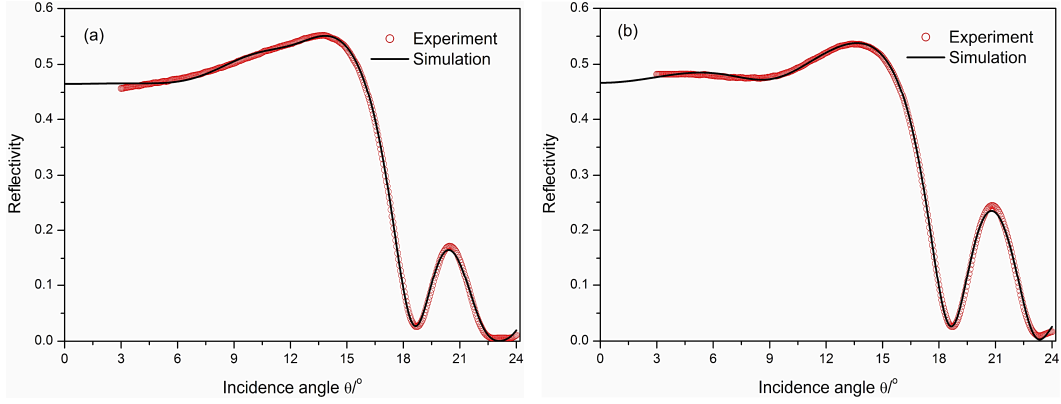} 
	\caption{Experimetal and fitted reflectivity of the fabricated broadband multilayer mirror as a function of the incidence angle for a wavelength of 13.5nm. (a) The Mo/Si multilayer is fabricated based on the designed thickness distribution of Robust Design 1 as shown in Fig. 1; (b) The designed Mo/Si multilayer structure of Robust Design 2 which is demonstrated in Fig. 1 and used in the fabrication.}
\end{figure}

The designed and fitted layer thickness distributions of Robust Design 1 and Robust Design 2 are shown in Fig. 4(a) and Fig. 4(b), respectively, and one can found the random differences between the designed and fitted layer thicknesses. It is also found that the random thickness flunctions in Fig. 4(b) are larger than that in Fig. 4(a), but the corrsponding reflectivity plateau of Robust Design 2 in Fig. 3(b) is analogous to that of Robust Design 1 in Fig. 3(a). This result can be understood that the thickness distribution of Robust Design 2 has a more stability of the reflectivity plateau with respect to random layer thickness fluctuations. Therefore, the advantage of robust multilayer design based on multiobjective evolutionary algorithm has been demonstrated in the experiments, and our scheme can be useful in reducing the production risks of EUV mirrors which are high cost. In this research, we only focused on the reflectivity profile and the sensitivity to random thickness flunctions of Mo/Si multilayer, while the future improvement of the designing procedure can be the development of multiobjective evolutionary algorithm to simultaneously optimize the reflectivity profile, robust peformance and relative dispersion \cite{Yakshin:2010,Kozhevnikov:2015}. Moreover, our method can also be extended to design and fabricate the robust multilayers for hard X-rays \cite{Pardini:2016}.

\begin{figure}
	\includegraphics[width=14cm]{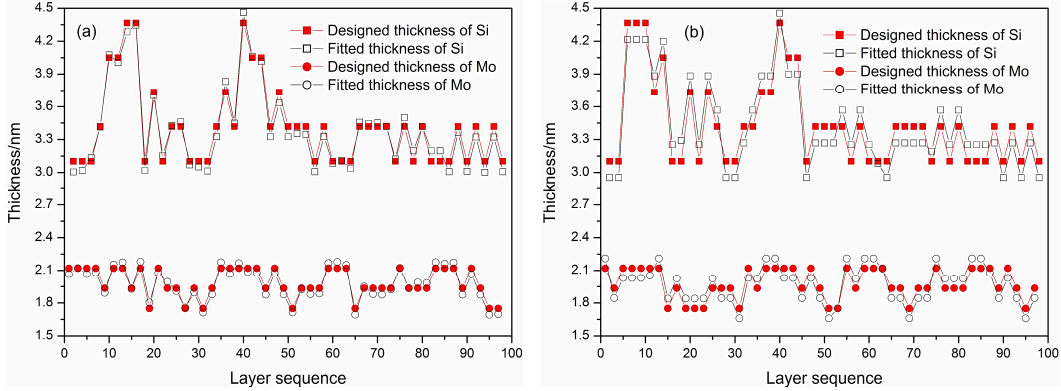}
	\caption{Designed layer thickness distribution of broadband Mo/Si multilayer and the fitting structure to the experimental reflectivity curve. (a) and (b) are the results of designed multilayer structures of Robust Design 1 and Robust Design 2, respectively. Here, the naturally formed interlayers are considered, but these interlayers are not presented in both graphs.}
\end{figure}   

\section{Conclusion}
The designing procedure of broadband Mo/Si multilayer based on multiobjective evolutioary algorithm is improved, and the robust multilayer structures which adapt the coating system's control precision of layer thickness are obtained. The designed EUV multilayers are fabricated and charaterized, and these analyses demonstrate the adavantages of robust multilayer design for broadband EUV multilayer mirror. Furthermore, this procedure can also be used in the robust design of suppermirrors in the X-ray range.  

\begin{acknowledgments}
This work was supported by National Natural Science Foundation of China (61405189) and
Jilin Scientific and Technological Development Plan (20150101019JC, 20170312024ZG).
\end{acknowledgments}

\end{document}